\documentclass[a4paper,11pt]{article}
\usepackage[margin=1in]{geometry}
\usepackage{amsmath}
\usepackage{mathrsfs}
\usepackage{amssymb}
\usepackage{simpler-wick}
\usepackage{dsfont}
\usepackage{hyperref}
\usepackage{graphicx}
\usepackage{pdfpages}
\usepackage{blindtext}
\usepackage{cite}
\usepackage{rotating}
\usepackage{fancyhdr}
\usepackage[capitalise]{cleveref}

\newcommand{\cmmnt}[1]{\ignorespaces}

\title{}

\begin{document}

\begin{titlepage}
\setcounter{page}{1} \baselineskip=15.5pt 
\thispagestyle{empty}

\begin{center}
{\fontsize{18}{18} \bf Rational Wavefunctions in de Sitter Spacetime\vspace{0.1cm}
\;}\\
\end{center}

\vskip 18pt

\begin{center}
\noindent
{\fontsize{12}{18}\selectfont Harry Goodhew\footnote{\tt hfg23@cam.ac.uk}}
\end{center}

\begin{center}
  \vskip 8pt
\textit{Department of Applied Mathematics and Theoretical Physics, University of Cambridge, Wilberforce Road, Cambridge, CB3 0WA, UK}
\abstract{The Bootstrap approach to calculating cosmological correlators relies on a well motivated ansatz. It is typical in the literature to assume that correlators are rational functions as this greatly increases our constraining power. However, this has only previously been demonstrated for some specific theories. In this paper we find a set of assumptions which we prove are sufficient to ensure that the wavefunction coefficients are rational. As a corollary of this we generalise the manifestly local test to higher dimensions. This result greatly reduces the allowed space of functions that wavefunction coefficients can take in both the Effective Field Theory of Inflation and Pure Gravity models and is thus a key ingredient in the Cosmological Bootstrap program. }
\end{center}

\end{titlepage}

\section{Introduction}\label{sec:Intro}
The history of the universe is encoded in the matter and radiation that it contains. Within the current leading paradigm of the early universe all deviations from homogeneity and isotropy today were sourced by quantum fluctuations of some, currently unknown, fields during an initial phase of approximately de Sitter expansion known as inflation. We have access to these fields through their statistical properties, which are imprinted onto everything we see in the universe today. This gives us a window through which we can learn about the physics of our universe at energies far higher than we can recreate on earth. So, understanding the statistics of these fields, in particular their correlation functions and the wavefunction from which they originate is a major goal of theoretical cosmology.

In recent years much progress has been made towards developing bootstrap techniques\cite{Creminelli:2011mw,Kehagias:2012pd,Mata:2012bx,Bzowski:2013sza,Ghosh:2014kba,Kundu:2014gxa,Kundu:2015xta,Pajer:2016ieg,Bzowski:2019kwd,Baumann:2019oyu,PSS,Isono:2020qew,Green:2020ebl,Baumann:2020dch,benincasa2022wavefunctionals} for both wavefunction coefficients and correlators. By now the combination of unitarity\cite{COT,sCOTt,Cespedes:2020xqq,goodhew2021cutting}, locality\cite{MLT} and the flat space amplitude limit\cite{Raju:2012zr,Maldacena:2011nz,Arkani-Hamed:2018kmz,benincasa2022amplitudes} have been successful in constraining the most general form of three point functions for both scalars\cite{BBBB} and tensors\cite{cabass2022bootstrapping,cabass2022graviton} as well as specific four point functions\cite{bonifacio2021amplitudes} and loop corrections to the power spectrum\cite{sCOTt}, for a review see\cite{Baumann:2022jpr}. However, to fix these functions at tree level we require a very restrictive, polynomial, ansatz\cite{BBBB,cabass2022bootstrapping}. This is analogous in scope and constraining power to the flat space result that tree level amplitudes can only contain at worst simple poles of invariants built from the exchanged 4-momenta\cite{Benincasa:2007xk,TASI}. A priori we cannot make this assumption and in fact we find logarithmic terms in, for example, non-derivative contact interactions involving both massless and conformally coupled fields\cite{falk1992angular,Arkani-Hamed:2018kmz}. If present, such divergences significantly weaken the constraining power of both the cosmological optical theorem\cite{COT} and related cutting rules\cite{sCOTt,goodhew2021cutting} as well as the manifestly local test\cite{MLT}. This is because there are many more ways to match the discontinuities and singularities they require if more general functions are allowed. 

In this paper we prove that, for a theory that is:
\begin{itemize}
    \item Massless,
    \item Bosonic,
    \item Scale invariant,
    \item In even spacetime dimension,
    \item Parity even,
    \item Built from interactions with at least two derivatives, and
    \item Tree level
\end{itemize}
all wavefunction coefficients are rational functions in the energies of the internal and external particles (plus perhaps some contractions of helicity tensors and three momenta). The conditions that are required on these fields may seem quite restrictive. However, in even spacetime dimension, they cover all gravitational interactions which are built from second derivatives of the metric through the Riemann tensor. Furthermore, they also are sufficient to capture the effective field theory (EFT) of single field inflation. This is because the EFT of inflation is a theory of a massless goldstone mode and so is built entirely from its derivatives in the scale invariant limit\cite{pich2018effective}. This work builds on and generalises the results in\cite{anninos2015late} where they show that the tree level wavefunction is free from logarithmic divergences for gauge fields and gravity in $3+1$ spacetime dimensions.

To do this we will start in \cref{sec:Free} by demonstrating that it is possible to express the solution to the equations of motion as a series containing no odd powers of the conformal time less than the number of spatial dimensions. As a corollary of this we will derive an extension to the manifestly local test\cite{MLT} that is valid in arbitrary spacetime dimension. We then show, in \cref{sec:Contact}, that this result is incompatible with logarithmic divergences from contact interactions of two derivative theories. This is done by first considering the time dependence arising from the time integrals. Then using an altered ansatz for the solution to demonstrate that the only possible irrational function of the momentum that can contribute must be accompanied by a logarithmic time divergence. In \cref{sec:Exchange}, we extend these arguments to exchange diagrams. In \cref{sec:Conclusion}, we first briefly discuss how breaking the assumptions outlined above invalidates these arguments and reintroduces the possibility of logarithmic divergences. Then, finally, we highlight some potential future directions for extending this work and make some concluding remarks.

\section{The Free Theory}\label{sec:Free}
In this section we find a series solution to the equations of motion which, for massless particles, contain no odd powers of $k$ or $\eta$ less than the spatial dimension. In order to make these discussions concrete whilst staying as theory agnostic as possible, we will consider a general class of free theories. In $d+1$ dimensional spacetime, for traceless\footnote{If we also wish to consider fields with a non-zero trace we can subtract the trace and treat it as an additional scalar field.}, integer spin fields we use the free action developed in \cite{bordin2018light} and discussed in the context of the bootstrap method in \cite{goodhew2021cutting},
\begin{equation}
S=\int d^{d+1}x a^{d-1}\frac{1}{2s!}\left[{\Phi_{i_1\dots i_s}'}^2-c_s^2\left(\partial_j\Phi_{i_1\dots i_s}\right)^2-\delta c_s^2 \left(\partial^j\Phi_{ji_2\dots i_s}\right)^2-m^2a^2\left(\Phi_{i_1\dots i_s}\right)^2\right].
\end{equation}
Here, $'$ indicates derivatives with respect to conformal time $\eta$ and the latin indices, $i,j$, span the $d$ dimensional spacelike hypersurface orthogonal to this coordinate. We have enforced scale invariance by including inverse factors of the scale factor for each coordinate derivative. Just as in \cite{goodhew2021cutting} we Fourier transform and diagonalise this using the helicity modes, $\Phi_h$, defined by
\begin{equation}
\Phi_{i_1\dots i_s}=\Phi_h \epsilon^h_{i_1\dots i_s}.
\end{equation}
These helicity tensors are defined as an outer product of helicity vectors,
\begin{equation}
\epsilon^h_{i_1\dots i_s}=\epsilon^{h_1}_{i_1}\dots \epsilon^{h_s}_{i_s},
\end{equation}
which are themselves defined so that they satisfy
\begin{align}
\epsilon^h_i(-\textbf{k})&=\left[\epsilon^h_i(\textbf{k})\right]^*,& \left[\epsilon_i^h(\textbf{k})\right]^*\epsilon_i^{h'}(\textbf{k})&=4\delta^{h h'}.
\end{align}
Note that these fields are not assumed to be transverse, $h$ is allowed to take $d$ different values including $0$ where $\epsilon^0$ is proportional to the momentum. The contributions from the other helicity modes are therefore transverse by the orthogonality condition. The equations of motion in terms of these variables are therefore
\begin{equation}
\eta^2 \Phi''_{h}+p(\eta)\eta \Phi'_{h}+q_{h}(\eta)\Phi_{h}=0,
\end{equation}
where
\begin{align}
p(\eta)&=1-d,&q_{h}(\eta)&=\frac{m^2}{H^2}+\eta^2 c_s^2k^2+\eta^2\delta c_s^2 k^2 \lambda_h,
\end{align}
and $k^2\lambda_h$ are the eigenvalues corresponding to each helicity mode $h$, the $k^2$ has been factored out so that $\lambda_h$ is just a number, independent of $k$ or $\eta$. From this point onwards we will be setting $H=1$ for notational simplicity. Due to the physical origin of each of the terms in $p$ and $q$ they are guaranteed to be analytic everywhere except, perhaps, in the infinite past. 

This equation is of the form studied by Frobenius\cite{boas1999mathematical} and therefore has a solution that converges everywhere in the range $[-\infty,0)$ of the form 
\begin{equation}
\Phi_h=\eta^\Delta\sum_{k=0}^\infty A_k \eta^k.
\end{equation}
Where $\Delta$ satisfies the indicial polynomial,
\begin{equation}
I(\Delta)=\Delta(\Delta-1)+(1-d)\Delta+m^2=0 \Rightarrow \Delta_\pm= \frac{d}{2}\pm\sqrt{\frac{d^2}{4}-m^2}=\frac{d}{2}\pm\nu.
\end{equation}
When $\Delta_+-\Delta_-=2\nu$ is not an integer we therefore have two linearly independent solutions that can be defined recursively  
\begin{align}
\Phi_h^\pm&=\eta^{\Delta_\pm}\sum_{n=0}^\infty A_n^{\pm}\eta^n,& A_n^{\pm}=\frac{-1}{I(n+\Delta_\pm)}\sum_{m=0}^{n-1}\frac{q_h^{(n-m)}(0)}{(n-m)!}A_m^{\pm}.
\end{align}
Where $A_0^{\pm}$ are fixed by the initial (or boundary) conditions. 

When $2\nu$ is an integer these two solutions are not guaranteed to be linearly independent and it is possible for there to be a logarithmic term. In these cases, the method of Frobenius tells us to take a pair of solutions of the form
\begin{align}
\Phi_h^1&=\Phi_h^+,\\
\Phi_h^2&=C\Phi_h^1\log(\eta)+\eta^{\Delta_-}\sum_{n=0}^\infty B_n\eta^n.
\end{align}
The coefficients $B_n$ and $C$ are fixed by the equations of motion,
\begin{equation}
\sum_{n=0}^\infty 2C\eta^{2\nu}A_n^{+}\eta^n(\nu+n)+\eta^nB_n\left((n+\Delta_-)(n-\Delta_+)+\sum_{m=0}^\infty \frac{q_h^{(m)}(0)}{m!}\eta^m\right)=0.
\end{equation}
For $n<2\nu$ this expression is exactly the same as the equation fixing $A_n^{h,-}$ and so $B_n=A_n^{h,-}$ for $n<2\nu$. However, for $n=2\nu$ we find
\begin{equation}
2CA_0^{+}\nu+B_{2\nu}(-\Delta_+\Delta_-+q_{h}^{(0)})+\sum_{m=1}^{2\nu}\frac{q_h^{(m)}(0)}{m!}A_{2\nu-m}^{-}=0.
\end{equation}
The term multiplying $B_{2\nu}$ is zero and as such we cannot constrain it so it can be set arbitrarily. In fact we can see that any non-zero contribution coming from $B_{2\nu}$ is degenerate with $\Phi^1_h$ so we can absorb this arbitrariness into $A_0$. Furthermore, for finite $A_{2\nu}$, the sum vanishes,
\begin{equation}
\sum_{m=1}^{2\nu}\frac{q_h^{(m)}(0)}{m!}A_{2\nu-m}^{-}=\sum_{m=0}^{2\nu-1}\frac{q_h^{(2\nu-m)}(0)}{(2\nu-m)!}A_{m}^{-}=-I(\Delta_+)A_{2\nu}^{-}=0,
\end{equation}
which implies that
\begin{equation}
2CA_0^{+}\nu=0.
\end{equation}
We know that $A_0^{+}\neq0$ as this would result in $\Phi_h^1=0$. Then, we must have either $C=0$ or $\nu=0$. Therefore, for $\nu\neq0$ and finite $A_{2\nu}$, we will be free from these logarithmic contributions to the modefunction and the solution can be written as a sum of powers of $\eta$. 

Scale invariance fixes $m,\ c_s$ and $\delta c_s$ to be constants and so all odd coefficients will vanish. To see this first note that the only the 0th and 2nd derivatives of $q$ are non-zero,
\begin{equation}
q_h^{(n)}=m^2\delta_{n0}+(c_s^2+\delta c_s^2 \lambda_h)k^2 \delta_{n2}.
\end{equation}
This gives us a closed form expression for the coefficients,
\begin{align}
A_{2n}^{\pm}&=A_0^{\pm}k^{2n}\left(\frac{c_h^2}{-4}\right)^n\frac{\Gamma(1\pm\nu)}{\Gamma(1+n)\Gamma(1\pm \nu+n)},\\A_{2n+1}^\pm&=0,
\end{align}
where $c_h^2=c_s^2+\delta c_s^2\lambda_h$. For integer $\nu$ we can see that this expression diverges for $2n=2\nu$ and so we cannot conclude that $C=0$. This can also be seen as insisting that $B_{2\nu}=0$ terminates this sum but the resulting polynomial is not a solution to the equations of motion. For odd integer $2\nu$ we can replace $A_n^-$ with $B_n$ for all $n$. Fixing $B_{2\nu}=0$ in this case ensures that the two are equal beyond $n=\nu$. The general solution to the equations of motion for non-integer $\nu$ can be written as a linear combination of these two sums,
\begin{equation}
\Phi_h=
B_0\sum_{n=0}^{\infty} \frac{ \Gamma(1-\nu)(c_hk\eta)^{2n+\Delta_-}}{(-4)^n\Gamma(1+n)\Gamma(1-\nu+n)}+A_0 \sum_{n=0}^\infty \frac{\Gamma(1+\nu)(c_hk\eta)^{2n+\Delta_+}}{(-4)^n\Gamma(1+n)\Gamma(1+\nu+n)}.
\end{equation}
Here we have redefined the arbitrary coefficients to include powers of $c_hk$ that will be convenient later and dropped the $+$ label. This expression is valid for all non-integer values of $\nu$ as we recovered these series solutions when $2\nu$ is an odd integer. The bulk-boundary propagator coming from this solution is
\begin{equation}\label{eq:K}
K_k^h(\eta)=\left(\frac{\eta}{\eta_0}\right)^{\Delta_-}\left(\sum_{n=0}^{\infty}\frac{ \Gamma(1-\nu)(c_hk\eta)^{2n}}{(-4)^n\Gamma(1+n)\Gamma(1-\nu+n)}+\frac{A_0}{B_0}\sum_{n=0}^\infty \frac{\Gamma(1+\nu)(c_hk\eta)^{2n+2\nu}}{(-4)^n\Gamma(1+n)\Gamma(1+\nu+n)} \right).
\end{equation}
The important point to note here is that, for massless particles, $\Delta_-=0$, there are no odd powers of $\eta$ less than $\eta^d$ in this expression. This will be the key observation that will allow us to exclude the possibility of logarithmic divergences\footnote{I would like to thank Enrico Pajer for sharing an unpublished manuscript suggesting this approach to the problem of identifying divergences.}. It is interesting that this absence of odd powers of $\eta$ is precisely what is guaranteed by the Fefferman-Graham expansion of the metric. This was used in \cite{anninos2015late} to conclude that gravity contains no logarithmic divergences. The formalism used here allows us to extend this result to higher (and lower) spins. The bulk-bulk propagator is
\begin{equation}
    G_p^h(\eta,\eta')=\begin{cases}\displaystyle
        \frac{\Phi_h^1(p,\eta')\Phi_h^2(p,\eta)}{(-\eta')^{1-d}W(\Phi_h^1,\Phi_h^2)}&\eta'\leq\eta,\\\displaystyle
        \frac{\Phi_h^1(p,\eta)\Phi_h^2(p,\eta')}{(-\eta')^{1-d}W(\Phi_h^1,\Phi_h^2)}&\eta\leq\eta',
    \end{cases}
\end{equation}
where $\Phi_h^{1/2}$ are solutions with coefficients chosen so that they satisfy the specified boundary conditions. The factor of $(-\eta')^{1-d}$ arises because the propagator is the Green's function for the equation
\begin{equation}
    a^{d-1}G''+p(\eta)a^{d}G'+q(\eta)a^{1+d}G=\delta(\eta-\eta'),
\end{equation}
so we must adjust the junction condition accordingly.

\subsection{The Manifestly Local Test}\label{sec:MLT}
We now make a brief aside to the main topic of the paper to consider the Manifestly Local Test (MLT) for more general theories.  Consider an arbitrary wavefunction coefficient. This is built out of bulk-bulk and bulk-boundary propagators plus some differential operators, $\mathcal{O}_a$ that act on them,
\begin{equation}
    \psi_N=\int \prod_{a}\frac{d\eta_a}{\eta_a^{d+1}} \prod_b \mathcal{O}_b K_{k_b}^{h_b} \prod_c \mathcal{O}_c G_{p_c}^{h_c}(\eta,\eta').
\end{equation}
The operators $\mathcal{O}_a$ are arbitrary except they cannot contain any inverse Laplacians (this is the sense in which the theory is ``Manifestly'' Local). For massless particles in $3+1$ spacetime dimensions the MLT\cite{MLT} constrains the second term in the Taylor expansion about zero of all such wavefunction coefficients,
\begin{equation}
    \left.\frac{\partial \psi_n}{\partial k_a}\right\rvert_{k_a=0}=0.
\end{equation}
The MLT may appear to some readers like a consistency relation. However, it is distinct as it holds away from physical momenta. 

This property is inherited from the fact that there is no term linear in $k$ in the bulk-boundary or bulk-bulk propagators. It is not spoiled by the presence of the operators $\mathcal{O}$ as momentum and time derivatives commute whilst spatial derivatives just bring down additional even powers of the momentum,
\begin{equation}
    \textbf{k}_a\cdot\textbf{k}_a=k_a^2,
\end{equation}
or
\begin{equation}
    \textbf{k}_a\cdot\textbf{k}_b=\frac{1}{2}\left[(\textbf{k}_a+\textbf{k}_b)\cdot(\textbf{k}_a+\textbf{k}_b)-k_a^2-k_b^2\right]=\frac{1}{2}\left(p^2-k_a^2-k_b^2\right).
\end{equation}
Note that the internal momenta, $p$, are regarded as independent variables when performing this derivative. There may also be some polarisation factors that contract with the momenta but these are explicitly stripped off in the application of the MLT and so we ignore them from the outset here. 

In the general dimensional case we find that the first odd power of $k$ in the bulk-boundary propagator is $k^{2\nu}$ and so
\begin{equation}\label{eq:MLT}
    \left.\frac{\partial^n K_{k}^{h}(\eta)}{\partial^n k}\right\rvert_{k=0}=0,\ \forall\textrm{ odd } n<2\nu.
\end{equation}
In particular, for massless fields this is true for all odd $n<d$. This follows straightforwardly from the series solution in \cref{eq:K}. Just as in the $d=3$ case, this property of the propagators is passed on to the wavefunction coefficients.  One might worry about the potential for $A_0$ and $B_0$ to depend on $k$. However, if we look at the equations of motion we can, by making the substitution $\eta\rightarrow x=c_hk\eta$, conclude that all $k$ dependence in the solution is through $x$. Therefore, the only additional $k$ dependence possible is through an overall scaling which is explicitly canceled in the bulk-boundary propagator by its late time limit. 

To see the power of this result consider the three point interactions of a massless field in $d=5$ with two derivatives. The wavefunction coefficient will have a total energy pole of order $3$\cite{COT,MLT}. The most general time independent contribution to the solution that we can write down with the correct scaling\footnote{Scale invariance demands that all wavefunction coefficients for massless fields scale like the momentum to the power of the dimension.} is thus
\begin{multline}
    k_T^3\psi_3=C_{00}k_T^8+C_{10}e_2k_T^6+C_{01}e_3k_T^5+C_{20}e_2^2k_T^4+C_{11}e_2e_3k_T^3+C_{30}e_2^3k_T^2+C_{02}e_3^2k_T^2\\+C_{21}e_2^2e_3k_T+C_{40}e_2^4+C_{12}e_2e_3^2,
\end{multline}
where $k_T,\ e_2,\ e_3$ are the elementary symmetric polynomials,
\begin{align}
    k_T&=k_1+k_2+k_3\\
    e_2&=k_1 k_2+k_1 k_3+k_2 k_3\\
    e_3&=k_1 k_2 k_3.
\end{align}

We expect to find two distinct wavefunction coefficients, arising from the interactions $(\partial_i\phi)^2\phi$ and ${\phi'}^2\phi$. However, the $d=3$ MLT is only enough to reduce the number of coefficients from $10$ to $5$. The second condition that this extended MLT gives us is exactly sufficient to further reduce the number of free coefficients to $2$,
\begin{multline}
    k_T^3\psi_3=\frac{1}{3}(C_{01}-5C_{00})e_2e_3^2+\frac{1}{2}(C_{01}-5C_{00})e_2^2e_3k_T+\frac{1}{6}(C_{01}-5C_{00})(3e_2^3-e_3^2)k_T^2\\-C_{01}e_2e_3k_T^3+\frac{1}{2}(15C_{00}-C_{01})e_2^2k_T^4+C_{01}e_3k_T^5-5C_{00}e_2k_T^6+C_{00}k_T^8.
\end{multline}
We recover results proportional to $(\partial_i\phi)^2\phi$ for $C_{00}=C_{01}$ and to ${\phi'}^2\phi$ for $C_{00}=0$. Note that we have ignored the possibility of time divergences in this ansatz. If we had included them we would have been able to constrain the allowed terms to 
\begin{equation}
    \psi_3\supset \frac{A(k_T^2-2e_2)}{\eta_0^3}+\frac{B(2e_3k_T-e_2^2)+C(k_T^2-2e_2)^2}{\eta_0}.
\end{equation}
These are precisely the divergent contributions to the two derivative interactions considered.

The absence of odd powers of $k$ less than $k^{2\nu}$ in the bulk-boundary propagator is true for all real $\nu$\footnote{The series expansion also holds for imaginary $\nu$ but this size comparison loses meaning in that case.}. Furthermore, for integer $\nu$ the logarithmic terms enter first at $(k\eta)^{2\nu}\log(k\eta)$ and all lower powers of $k$ are even. This formalism thus further demonstrates that the MLT extends beyond massless fields and holds for all sufficiently light fields. We can see this in $d=3$ for the case of two conformally coupled fields interacting with a massive field which has a three point function given by\cite{Arkani-Hamed:2015bza,COT}
\begin{equation}
    \psi_3^{\varphi\varphi\sigma}(k_1,k_2,k_3)\propto k_3^{-\frac{1}{2}+\nu}\left._2F_1\right.\left[\frac{1}{2}-\nu,\frac{1}{2}+\nu,1,\frac{k_3-k_1-k_2}{2k_3}\right].
\end{equation}
Here $k_1$ and $k_2$ are the momenta of the conformally coupled fields and $k_3$ is the momentum of the massive field whose mass gives us $\nu$. 
Conformally coupled scalars have $2v=1$ and so we have no expectation that the first derivative with respect to $k_1$ or $k_2$ will vanish. However, the first derivative with respect to $k_3$ is
\begin{equation}
    \lim_{k_3\rightarrow 0}\partial_{k_3}\psi_3^{\varphi\varphi\sigma}(k_1,k_2,k_3)\propto \lim_{k_3\rightarrow 0}k_3^{2\nu-1}(k_1+k_2)^{-\frac{1}{2}-\nu}+\mathcal{O}(k_3),
\end{equation}
which vanishes for $2\nu>1$ exactly as expected from \cref{eq:MLT}.
\section{Contact Diagrams}\label{sec:Contact}
In this section we demonstrate that for theories satisfying the assumptions in \cref{sec:Intro} all contact interactions generate wavefunction coefficients that are rational functions of the energy. For theories involving gravity some of these terms will come from solving the constraint equations which are, generically, differential equations relating the transverse, traceless component of the metric to its other components on the final slice of inflation. This is potentially problematic as it drastically increases the complexity of the allowed interaction terms. However, if we choose to decompose the metric in the ADM formalism in the unitary gauge\cite{arnowitt2008republication,Maldacena2003} the lapse and shift enter into these equations with no time derivatives. So, in Fourier space, these equations become non-dymamical and can be solved algebraically. We are, therefore, in a specific gauge but the absence of divergences in the correlation functions is a physical statement and so must be gauge independent. This does not completely remove the complication however, as it requires us to take into consideration the possibility of inverse spatial derivatives. One might worry that this spoils the assumption that our theory contains interactions that involve at least two derivatives. Fortunately, the net number of derivatives on all the interaction terms (counting these inverse derivatives negatively) entering the equations will remain at least two. This is ensured by the dimensionality of the constraint equations.

To begin with we consider only contact diagrams. We will assume throughout that we have removed all time derivatives higher than the first using the equations of motion to arbitrarily high order in perturbation theory. This avoids the complications that higher derivatives bring to the definition of the conjugate momentum. For massive fields the equations of motion allow us to replace a term containing a second time derivative with the mass term which has no derivatives. Therefore, such theories will generically contain interactions involving no derivatives. This will spoil our arguments and so we must at this point restrict ourselves to massless fields. In odd spacetime dimension $2\nu=d$ is even and so we expect logarithms in the propagator and, by extension, potentially in the correlators. In even spacetime dimension, however, the propagator is given by
\begin{equation}\label{eq:propK}
K_k^h(\eta)=\left(\sum_{n=0}^{\infty}\frac{ \Gamma\left(1-\frac{d}{2}\right)(c_hk\eta)^{2n}}{(-4)^n\Gamma(1+n)\Gamma\left(1-\frac{d}{2}+n\right)}+\frac{A_0}{B_0}\sum_{n=0}^\infty \frac{\Gamma\left(1+\frac{d}{2}\right)(c_hk\eta)^{2n+d}}{(-4)^n\Gamma(1+n)\Gamma\left(1+\frac{d}{2}+n\right)} \right).
\end{equation}

Even for massless fields we are able to reduce the number of time derivatives by one as the equations of motion contain both a first and a second time derivative. In the scale invariant limit of the effective field theory of single-clock inflation, all interaction terms are built from derivatives of fields and so there are always at least three fields that have derivatives acting on them (as interaction terms involve at least three fields multiplying each other). Because of this, the equations of motion cannot reduce the number of derivatives to zero and so, there will always be at least three derivatives acting on the fields even after removing all higher time derivatives. For multifield inflation this is not guaranteed as the other fields can have non-derivative couplings with the inflaton. However, we will assume that these interactions between fields also contain at least two derivatives and can be reduced to single time derivatives to arbitrary order in perturbation theory. 

Likewise, in the case of the gravity, as was discussed in\cite{cabass2022graviton}, the corrections to the action are constructed from derivatives of the perturbation to the extrinsic curvature and Riemann curvature tensor on the hypersurface at the end of inflation. These perturbations to the extrinsic curvature are built from first time derivatives of the metric whilst the Riemann curvature tensor is built out of derivatives and products of the Christoffel symbols, which each contain first spatial derivatives of the metric. Therefore, any rotation invariant combinations of the two will contain at least first derivatives on a minimum of two terms in the interaction. This ensures that it is not possible to reduce any such interaction terms to ones including fewer than two derivatives. One could, in principle, further reduce the number of derivatives acting on individual terms through integration by parts at the level of the action. However, we will avoid doing this as it produces needless complications due to boundary terms and the cancellation of divergences between the terms that result from the integration by parts.

\subsection{The Absence of Logarithms in Time}\label{sec:Time}
In this section we argue that contact correlation functions in theories satisfying the assumptions in \cref{sec:Intro} cannot have any logarithmic divergences in time. This will allow us to prove that there can be no irrational functions of momenta in the wavefunction coefficients in the following section. To start with, consider a parity even theory with only spatial derivatives. The contribution to the wavefunction coefficient from an interaction with some number, $2m$,\footnote{Note that we have now set the mass to zero and here $m$ is just some arbitrary positive integer} of spatial derivatives acting on $N$ fields is
\begin{equation}
\Psi_N\supset\lim_{\eta_0\rightarrow 0}\int_{-\infty}^{\eta_0}d\eta\frac{1}{\eta^{d+1}}\eta^{2m}F(\textbf{k}_a)\prod_a^N K_{k_a}^{h_a}(\eta).
\end{equation}
The function $F(\textbf{k}_a)$ is constructed from the appropriate contractions of the $N$ polarisation tensors and  $2m$ momenta. We allow each of the fields present to be distinct i.e. have a different sound speed or helicity state but all fields must be massless for this analysis to hold. 

When $d$ is even we already have logarithmic terms in the propagator and so generically expect logarithmic divergences, this case will not be considered further. However, for odd $d$, the only way to generate a logarithmic divergence from this integral is for it to contain a $\eta^{-1}$ term\footnote{This assumes that we can exchange the sum with the integral, which we will prove in \cref{sec:Momentum}.}. This requires the product over mode functions to contain an $\eta^{d-2m}$ term. The first odd power of $\eta$ in this product is $\eta^d$. Therefore, we cannot have an $\eta^{d-2m}$ term for any integer $m>0$ and there will not be a logarithm. A further important point here is that for parity odd interactions this analysis breaks down and it is once again possible to generate logarithmic divergences in even spacetime dimensions. Unless there are more derivatives than there are dimensions\cite{cabass2022bootstrapping} in which case all powers of $\eta$ are positive.

We also do not generate logarithms in theories with at least two time derivatives. To see this we replace some of the propagators with their first time derivative,
\begin{align}\nonumber
\frac{\partial_\eta K_k(\eta)}{a(\eta)}&=-\left(\sum_{n=1}^{\infty}\frac{ \Gamma\left(1-\frac{d}{2}\right)2n(c_hk\eta)^{2n}}{(-4)^n\Gamma(1+n)\Gamma\left(1-\frac{d}{2}+n\right)}+\frac{A_0}{B_0}\sum_{n=0}^\infty \frac{\Gamma\left(1+\frac{d}{2}\right)(2n+d)(c_hk\eta)^{2n+d}}{(-4)^n\Gamma(1+n)\Gamma\left(1+\frac{d}{2}+n\right)} \right)\\\label{eq:Kdot}
&=\eta^2\mathcal{K}_k(\eta).
\end{align}
The factor of $a$ in the denominator is included to ensure scale invariance. Notice that factorising out $\eta^2$ leaves $\mathcal{K}_k$ regular in the late time limit but changes the lowest odd power of $\eta$ to $d-2$. The wavefunction for an interaction with $n$ time derivatives is then
\begin{equation}\label{eq:timeonly}
\Psi_N\supset\lim_{\eta_0\rightarrow 0}\int_{-\infty}^{\eta_0}d\eta\frac{1}{\eta^{d+1}}\eta^{2n}\prod_i^n\mathcal{K}_{k_i}(\eta)\prod_j^N K_{k_j}(\eta).
\end{equation}
Note that, unlike in the case of spatial derivatives, $n$ and not $2n$ is the number of time derivatives. The factor of $2$ arises here due to the presence of the $\eta^2$ term in \cref{eq:Kdot}. In even spacetime dimensions, the lowest odd power of $\eta$ that is present in the product over the mode functions is now $d-2$ rather than $d$. This means that if we want to avoid logarithmic divergences we now need $d-2>d-2n$ and so we are guaranteed to avoid such terms provided $n>1$. 

Finally, we show that interactions with both $2m$ spatial derivatives and $n$ time derivatives cannot have logarithmic divergences. The spatial derivatives introduce a factor of $\eta^{2m}F(\textbf{k}_a)$ to \cref{eq:timeonly} which changes the above condition to $m+n>1$. We can combine the two cases with and without time derivatives to give a single condition which guarantees that we have no logarithmic divergences arising from parity even contact interactions in even spacetime dimensions,
\begin{equation}
    2m+n\geq 2,
\end{equation}
i.e. any theory with at least two derivatives. 

In deriving this result we have been completely agnostic to the possibility of other divergences in time. This is because the only other time divergences allowed by this ansatz are polynomial and such divergences are consistent with rational wavefunction coefficients. It also appears that we have been similarly agnostic to the initial conditions. However, insisting on the convergence of the integral (required for exchanging the sum and integral) turns out to fix early time behaviour, which can be understood as an initial condition.
\subsection{Divergences in Momenta}\label{sec:Momentum}
In this section we explore the potential divergences in the momentum to prove that the absence of logarithmic divergences in time ensures the rationality of the wavefunction. After performing the integrals in time there remains a time independent piece that could, theoretically, contain a term that depends logarithmically on the energies. As an example of such an integral consider,
\begin{equation}
\int_{-\infty}^0 \frac{e^{ik_1\eta}-e^{ik_2\eta}}{\eta}d\eta=\log(-ik_1)-\log(-ik_2).
\end{equation}

The series solution cannot say anything about the time independent contributions to the integral. So, it is necessary to consider a different ansatz to complete the proof that all contact diagrams will be rational functions of the energy,
\begin{equation}
    \Phi_h=e^{i\kappa\eta}P(\eta),
\end{equation}
where $P$ satisfies
\begin{equation}
    \eta^2P''+\eta(2i\kappa\eta -d+1 )P'+(i(1-d)\kappa\eta-\kappa^2\eta^2+c_s^2k^2\eta^2+\lambda_hk^2\delta c_s^2\eta^2)P=0.
\end{equation}
We then choose $\kappa^2=c_s^2k^2+\lambda_hk^2\delta c_s^2=c_h^2k^2$ to cancel the $\eta^2$ terms which reduces the equation to the form
\begin{equation}
    \eta^2P''+\eta(2ic_h k\eta -d+1 )P'+(i(1-d)c_h k\eta)P=0.
\end{equation}
This is an equation to which we can, once again, apply the method of Frobenius, 
\begin{equation}
    \eta^2P''+\eta\tilde{p}P'+\tilde{q}P=0.
\end{equation}
Moreover, the indicial equation is unchanged as $p(0)=\tilde{p}(0)$ and $q(0)=\tilde{q}(0)$ and so our two solutions are
\begin{align}
    P_1&=\eta^d\sum_{n=0}^\infty a_n\eta^n\,,\\
    P_2&=CP_1\log(\eta)+\sum_{n=0}^\infty b_n\eta^n\,.
\end{align}
Restricting to even spacetime dimension, we must have $C=0$ for consistency with the previous results and
\begin{align}
    a_n&=-\frac{1}{n(n+d)}\sum_{m=1}^{n-1}\frac{(m+d)\tilde{p}^{(n-m)}(0)+\tilde{q}^{(n-m)}(0)}{(n-m)!}a_m\\&=-\frac{2n-1+d}{n(n+d)}ic_h ka_{n-1},\\
    b_n&=-\frac{1}{n(n-d)}\sum_{m=0}^{n-1}\frac{mp^{(n-m)}(0)+q^{(n-m)}(0)}{(n-m)!}b_m\\&=-\frac{2n-1-d}{n(n-d)}ic_h k b_{n-1}.
\end{align}
We can see from this that, in even spacetime dimension, $b_{\frac{1+d}{2}}=0$ and so this solution is just a polynomial. Unlike the previous instance where we found a polynomial solution, here the equations tell us that the series terminates and so this is a solution. Rather than choosing our two solutions to be $P_1$ and $P_2$ as found here, we instead take $P_2$ and then use the fact that our initial differential equation for $\Phi_h$ was real to guarantee that if $\Phi_h$ is a solution then so too is $\Phi_h^*$ to give a second, linearly independent solution\footnote{The independence of these two solutions is not guaranteed but in this case was checked by calculating their Wronskian which is non-zero, as required for linear independence.}
\begin{align}
    \Phi_h^\pm=e^{\pm ic_h k \eta} \sum_{n=0}^{\frac{d-1}{2}}\frac{\Gamma\left(\frac{1-d}{2}+n\right)\Gamma\left(1-d\right)}{n!\Gamma\left(\frac{1-d}{2}\right)\Gamma\left(1-d+n\right)}(\mp 2ic_hk \eta)^n.
\end{align}

Acting on either of these solutions with some derivative in space or time we will bring down some factors of the momentum but will not change the order of the polynomial,
\begin{equation}
    \partial_\eta \Phi_h^+=ic_h k e^{ic_h k\eta}\sum_{n=0}^{\frac{d-1}{2}}\frac{\Gamma\left(\frac{1-d}{2}+n\right)\Gamma\left(1-d\right)}{n!\Gamma\left(\frac{1-d}{2}\right)\Gamma\left(1-d+n\right)}\left((- 2ic_h k \eta)^n-2n(-2ic_h k\eta)^{n-1}\right).
\end{equation}
The $N$ particle wavefunction coefficient generated from just the solutions with positive energy can be written as
\begin{equation}\label{eq:PProduct}
    \psi_N=\lim_{\substack{L\rightarrow \infty\\\eta_0\rightarrow 0}}\int_{-L}^{\eta_0} d\eta \frac{\eta^{n+2m}}{\eta^{d+1}}F(\textbf{k}_a)\prod_{a=1}^N \tilde{P}_{k_a}(\eta) e^{ic_T k_T \eta},
\end{equation}
where $c_Tk_T=\sum_{a=1}^N c_{h_i}k_i$ is the total energy and, just as before, $n$ is the number of time derivatives whilst $2m$ is the number of spatial derivatives. The polynomials $\tilde{P}_k(\eta)$ can be either $P_k(\eta)$ or its first time derivative as the precise coefficients of the polynomial will not be required for the following arguments. This product of polynomials will generate a new polynomial whilst the time dependent prefactor may generate some negative powers of $\eta$ multiplying the exponential. 

The computation of this integral reduces to several integrals of the form
\begin{equation}
    \lim_{\substack{L\rightarrow \infty\\\eta_0\rightarrow 0}} \int_{-L}^{\eta_0}d\eta \eta^n e^{ik \eta}.
\end{equation}
Which is just the incomplete gamma function. The limit as we take $L\rightarrow \infty$ is poorly defined for real $k$ and so we must impose a boundary condition. The most standard boundary condition, which we assume here, is that we start in the so called ``Bunch-Davies'' vacuum\cite{bunch1978quantum}. This is achieved by rotating our time coordinate at infinity so that there is a brief period of Euclidean time evolution which causes the integral to converge. It also requires us to only consider positive $k$ and so all external lines must be represented by the positive energy solution $\Phi_h^+$. This, likewise, fixes one of the coefficients in our method of Frobenius expansion in terms of the other but the details of how that is done will not be relevant to this discussion. 

Ensuring that this integral converges in the infinite past is sufficient to allow us to exchange the infinite sum in our series expansion with the integration. We don't need the integral to be finite in the limit $\eta_0\rightarrow 0$ because there are only finitely many terms in the sum that diverge in this limit. We can therefore separate the sum into terms that might diverge in this limit and the remaining terms that don't. The terms that don't are well behaved in the infinite past (after imposing the correct boundary condition) and so the sum can be taken outside of the integral. 

We now explore the $\eta_0\rightarrow 0$ limit. For $n\geq0$ this limit just gives a constant. However, for negative powers of $\eta$ we have
\begin{equation}\label{eq:Gamman}
    \lim_{\eta_0\rightarrow 0} \int_{-\infty(1-i\epsilon)}^{\eta_0}d\eta \eta^{-n} e^{ik \eta}=\lim_{\eta_0\rightarrow 0}\frac{e^{ik\eta_0}}{\eta_0^{n-1}}\sum_{m=0}^{n-2}C_m (ik\eta_0)^m+\frac{(ik)^{n-1}}{(n-1)!}\left(Ei(ik\eta_0)+i\pi\right).
\end{equation}
Where $Ei$ is the exponential integral which contributes a logarithmic divergence in this limit. Therefore, for consistency with the previous observation that the integral doesn't contain any logarithmic divergences, the coefficients of the polynomial must exactly conspire to cancel any of these exponential integral terms (plus the $i\pi$). Furthermore, these logarithmic divergences are the only non-rational contributions to this integral. Their absence therefore ensures that the final wavefunction coefficient will be a rational function of the energies. 

To see this cancellation in a simple case consider the three point wavefunction coefficient coming from an interaction with two spatial derivatives in $3+1$ dimensional spacetime,
\begin{equation}\label{eq:psi3}
    \psi_3=i \int_{-\infty}^{\eta_0}d\eta\textbf{k}_1\cdot\textbf{k}_2\left(\frac{1}{\eta^2}-i\frac{k_T}{\eta}-e_2+ie_3\eta\right)e^{ik_T\eta}+\textrm{ perms}.
\end{equation}
We can evaluate this integral exactly,
\begin{equation}
    \psi_3=\textbf{k}_1\cdot\textbf{k}_2\frac{i}{k_T^2 \eta_0}e^{ik_T\eta_0}(-k_T^2+i(k_T e_2+e_3)\eta_0+k_T e_3\eta_0^2)+\textrm{ perms}.
\end{equation}
If we had just considered the first term in the bracket in \cref{eq:psi3} we would have found
\begin{equation}
    \psi_3\supset -\frac{i}{\eta_0}e^{ik_T\eta_0}-k_T\left(Ei(i k_T \eta_0)+i\pi\right).
\end{equation}
The resulting exponential integral cancels with an identical expression coming from the second term in the bracket in \cref{eq:psi3} to produce the final answer above which is free from logarithmic divergences. We are therefore left with a rational function in the momenta plus, perhaps some terms that diverge polynomially in time. 

The rational nature of this result also ensures the absence of logs in a related class of integrals that will be important later which take the form
\begin{equation}\label{eq:ConjugateP}
    \int \frac{d\eta}{\eta^{d-1}}P^*_{k_1}(\eta)\prod_{a=2}^NP_{k_a}(\eta)e^{i\left(\sum_{a=2}^N c_{h_a}k_a-c_{h_1}k_1\right)\eta},
\end{equation}
provided the exponent is positive. This is because this integral can be found by switching the sign of $k_1$ in the previous result and changing the sign of a term in a polynomial will not generate logs.


\section{Exchange Diagrams}\label{sec:Exchange}
We now extend the arguments in the previous section to tree level exchange diagrams. The calculation of such diagrams is significantly more complicated due to the presence nested time integrals but, as we show, the result will always be a rational function of the energies. It is important to emphasise at this point that, although we allow for interactions between different fields, each of these fields must be massless for our arguments to apply. We do not consider the possibility of the exchange of massive particles and the extension to include such interactions reintroduces the possibility of logarithmic divergences. 

To show the absence of logarithms in time we once again return to the Frobenius ansatz. As we established when looking at the exponential ansatz we need to specify an initial condition in order for our integrals to converge in the infinite past which fixes $A_0=a_0B_0$,\footnote{The linearity of this condition follows from the linear equations of motion}
\begin{equation}
    \Phi_h^+=B_0\sum_{n=0}^{\infty}\frac{\Gamma\left(1-\frac{d}{2}\right)(c_hk\eta)^{2n}}{(-4)^n\Gamma(1+n)\Gamma\left(1-\frac{d}{2}+n\right)}+a_0B_0\sum_{n=0}^{\infty}\frac{\Gamma\left(1+\frac{d}{2}\right) (c_hk\eta)^{2n+d}}{(-4)^n\Gamma(1+n)\Gamma\left(1+\frac{d}{2}+n\right)} .
\end{equation}
Note, that we must now enforce all the assumptions in \cref{sec:Intro}. If we cannot exclude logarithmic divergences in the contact diagram then we expect logarithms (or worse) in the exchange diagrams. Indeed, this is exactly what is seen in conformally coupled theories in $d=3$ where the non-derivative four point single exchange diagram contains di-logarithms\cite{Arkani-Hamed:2015bza,Hillman:2019wgh}. The other, linearly independent, solution is its complex conjugate,
\begin{equation}
    \Phi_h^-=B_0^*\sum_{n=0}^{\infty}\frac{\Gamma\left(1-\frac{d}{2}\right)(c_hk\eta)^{2n}}{(-4)^n\Gamma(1+n)\Gamma\left(1-\frac{d}{2}+n\right)}+a_0^*B_0^*\sum_{n=0}^{\infty}\frac{\Gamma\left(1+\frac{d}{2}\right) (c_hk\eta)^{2n+d}}{(-4)^n\Gamma(1+n)\Gamma\left(1+\frac{d}{2}+n\right)}.
\end{equation}
We need the Green's function to vanish in both the infinite past and at the end of inflation and so we take the two solutions
\begin{align}
    \lim_{\eta\rightarrow -\infty(1-i\epsilon)}\Phi^1_h(\eta)=0&\Rightarrow\Phi^1_h=\Phi_h^+,\\\lim_{\eta\rightarrow 0}\Phi^2_h(\eta)=0&\Rightarrow\Phi^2_h=\Phi_h^--\frac{B_0^*}{B_0}\Phi_h^+.
\end{align}
The Wronskian of these solutions is
\begin{equation}
    W(\Phi_h^1,\Phi_h^2)=W(\Phi_h^+,\Phi_h^-)=c_hk(c_hk\eta)^{d-1}dB_0^*B_0(a_0^*-a_0).
\end{equation}
So, the appropriate Green's function is
\begin{equation}\label{eq:PolyGreen}
    G_p^h(\eta,\eta')=\frac{{\eta'}^{d}}{d}K_p^h(\eta)\sum_{n=0}^{\infty}\frac{\Gamma\left(1+\frac{d}{2}\right)(c_hp\eta')^{2n}}{(-4)^n\Gamma(1+n)\Gamma\left(1+\frac{d}{2}+n\right)}\theta(\eta'-\eta)+\eta\leftrightarrow \eta'.
\end{equation}
The contribution from a general exchange diagram with $I$ internal lines to the $N$ point wavefunction coefficient is
\begin{equation}
    \Psi_N\supset\lim_{\eta_0\rightarrow0}\int_{-\infty}^{\eta_0} \prod_a^{I+1}\frac{d\eta_a}{\eta_a^{d+1}}\eta_a^{2m_a}F_a(\{\textbf{k}\}_a)\prod_b^N K_{k_b}^{h_b}(\eta)\prod_c^IG_{p_c}^{h_c}(\eta,\eta')\,,
\end{equation}
where $\eta$ and $\eta'$ are arbitrary times in the set $\{\eta_a\}$, $2m_a$ is the number of spatial derivatives acting at the vertex $\eta_a$ and  $\{\textbf{k}\}_a$ is the set of momenta entering the vertex at $\eta_a$. We don't consider time derivatives explicitly here but in an identical way to before these arguments will generalise to terms with time derivatives on the bulk-boundary propagators. We separately consider the case of derivatives on the bulk-bulk propagator as this can lead to an additional complication. 

It is always possible\cite{COT} to choose a vertex that is connected to only one other vertex. We will label this vertex $\eta$ and the vertex it is attached to $\eta'$. Isolating this term leaves us with an integral of the form
\begin{align}
    I&=\lim_{\eta_0\rightarrow0}\int_{-\infty}^{\eta_0} \frac{d\eta}{\eta^{d+1-2m}}F(\textbf{k}_a)\prod_a^{N_\eta}K_{k_a}^{h_a}(\eta)G_{p}^h(\eta,\eta')\\\nonumber&=
    \lim_{\eta_0\rightarrow0}\int_{\eta'}^{\eta_0} d\eta\eta^{2m-1}F(\textbf{k}_a)\prod_a^{N_\eta}K_{k_a}^{h_a}(\eta)\frac{1}{d}K_p^h(\eta')\sum_{n=0}^{\infty}\frac{\Gamma\left(1+\frac{d}{2}\right)(c_hp\eta)^{2n}}{(-4)^n\Gamma(1+n)\Gamma\left(1+\frac{d}{2}+n\right)}\\\label{eq:exchange}&+\int_{-\infty}^{\eta'} \frac{d\eta}{\eta^{d+1-2m}}F(\textbf{k}_a)\prod_a^{N_\eta}K_{k_a}^{h_a}(\eta)K_p^h(\eta)\frac{{\eta'}^{d}}{d}\sum_{n=0}^{\infty}\frac{\Gamma\left(1+\frac{d}{2}\right)(c_hp\eta')^{2n}}{(-4)^n\Gamma(1+n)\Gamma\left(1+\frac{d}{2}+n\right)}.
\end{align}

This integral, like $K$, contains no logarithmic divergences nor odd powers of $\eta$ less than $\eta^d$. This property allows us to replace any of the $K$'s in this integral with $I$ and conclude that the resulting integral will be free from such terms too. Repeating this for all the vertices in the diagram we can therefore conclude that we can never generate logarithmic divergences in this way. The absence of these small powers of $\eta$ can be seen in each term individually. To start with consider the final line. This is an integral of exactly the same form as we covered for the contact case and so we are guaranteed that the integral will vanish in the infinite past. Furthermore, at the upper limit any logarithmically divergent terms will vanish and we will just be left with a polynomial expression in $k$ with potentially some poles in $\eta'$. All such poles come from integrating even powers of $\eta$ and so will all be odd negative powers of $\eta'$. None of these terms can diverge faster than ${\eta'}^{2m-d}$ with $m\geq1$ and they are all multiplied by $\eta'^d$. Therefore, this final line can only contain positive powers of $\eta'$ and any powers of $\eta'$ less than ${\eta'}^d$ will be even. This is precisely the condition that prevented the generation of logarithmic divergences in the contact case. Therefore, we can guarantee that performing additional integrals over $\eta'$ will not contribute any logarithmic divergences. 

The first term is even more straightforward to deal with, it contains no negative powers of $\eta$ and, as $2m-1$ is odd, all the powers of $\eta$ less than $\eta^d$ in the integrand will be odd. Integrating such terms will only produce even powers of $\eta'$ which ensures that subsequent integrals cannot generate logarithmic divergences. Having removed this singly connected vertex we are left with a new diagram which must also have at least one singly connected vertex. All removed vertices contribute factors like $I$ which preserve all the relevant properties of $K$. Therefore, we can repeat this procedure for each vertex in the diagram and we will never generate a logarithmic divergence in time. 

As was mentioned previously, allowing single time derivatives to act on the bulk-boundary propagators will not alter this conclusion for the same reasons as in the contact case. However, we have not allowed for the possibility of time derivatives on the bulk-bulk propagator. It may seem straightforward to account for this by integrating the expression by parts to remove any time derivatives on singly connected vertices. This is true when there are no other time derivative terms but if there are other propagators with time derivatives this introduces second derivatives. Removing such terms with the equations of motion is the natural solution to this that we employed previously. This is only problematic in the instance that there is exactly one other term with a time derivative. If there is more than one other time derivative (or some spatial derivatives) then after application of the equations of motion we will still have at least two time derivatives and we return to familiar territory. However, if there is only one other time derivative then we have
\begin{align}\label{eq:Gderiv}
    I&=\lim_{\eta_0\rightarrow0}\int_{-\infty}^{\eta_0} \frac{d\eta}{\eta^{d-1}}\partial_\eta K_{k_1}^{h_1}(\eta)\prod_{a=2}^{N_\eta}K_{k_a}^{h_a}(\eta)\partial_{\eta}G_{p}^h(\eta,\eta')\\\nonumber&=\lim_{\eta_0\rightarrow0}\left[\frac{1}{\eta^{d-1}}\partial_\eta K_{k_1}^{h_1}(\eta)G_p^h(\eta,\eta')\right]_{-\infty}^{\eta_0}-\int_{-\infty}^{\eta_0} \frac{d\eta}{\eta^{d-1}}\partial_\eta K_{k_1}^{h_1}(\eta)\partial_\eta\prod_{a=2}^{N_\eta}K_{k_a}^{h_a}(\eta)G_{p}^h(\eta,\eta')\\&-\int_{-\infty}^{\eta_0} \frac{d\eta}{\eta^{d}}\left((1-d)\partial_\eta K_{k_1}^{h_1}(\eta)+\eta\partial^2_\eta K_{k_1}^{h_1}(\eta)\right)\prod_{a=2}^{N_\eta}K_{k_a}^{h_a}(\eta)G_{p}^h(\eta,\eta').
\end{align}
The boundary term vanishes due to the powers of $\eta$ in the $\eta\rightarrow0$ limit of $G$. The second term on this middle line contains two derivatives acting on different bulk-boundary propagators and so can be understood using the previous arguments. The final line contains only a single term with a time derivative which violates the two derivative condition we set earlier. Fortunately, the equations of motion relate this specific combination to the second spatial derivative and so
\begin{align}\nonumber
    I&=\int_{-\infty}^{\eta_0} \frac{d\eta}{\eta^{d-1}}\left(\left(c_{h_1}k_1\right)^2\prod_{a=1}^{N_\eta}K_{k_a}^{h_a}(\eta)-\partial_\eta K_{k_1}^{h_1}(\eta)\partial_\eta\prod_{a=2}^{N_\eta}K_{k_a}^{h_a}(\eta)\right)G_{p}^h(\eta,\eta').
\end{align}
Therefore, all terms with time derivatives acting on the bulk-bulk propagator can be integrated by parts to remove this time derivative in such a way that the remaining terms all contain at least two derivatives with no time derivatives higher than one. Thus, we were justified in ignoring the posibility of time derivatives on the propagator.

Just as in the contact case we haven't fully removed the possibility of logarithmic divergences in the momentum, only in time. However, recall the exponential ansatz for which the Green's function with internal momentum $p$ is
\begin{equation}\label{eq:Green}
    G_p(\eta,\eta')=\frac{iP_p(\eta)P^*_p(\eta')}{2  p^d (d-2)!!^2}e^{ic_hp (\eta-\eta')}\theta(\eta'-\eta)+\eta\leftrightarrow\eta'-\frac{iP_p(\eta)P_p(\eta')}{2 p^d (d-2)!!^2}e^{ic_h p(\eta+\eta')},
\end{equation}
where $n!!=n(n-2)\dots1$ for odd $n$. We can see from this expression that when we consider a singly connected vertex the necessary integral will be of the form \cref{eq:PProduct} or \cref{eq:ConjugateP} and we have already established that it will be absent any logarithmic divergences. Moreover, the resulting expression will still be of the form of a polynomial multiplied by an exponential, potentially with some negative powers of $\eta$. However, we know that no integral in the nested integral can contribute a logarithmic divergence in time. So, there is no way to generate an exponential integral term and all such contributions must necessarily cancel. 

One may worry that the exponent has the wrong sign in theories that have multiple speeds of sound\footnote{For equal sound speeds we can guarantee that the exponent has the correct sign because $\textbf{p}=\sum_{a}^{N_\eta}\textbf{k}_a$ and so 
\begin{equation}
    p^2=\left(\sum_{a}^{N_\eta}\textbf{k}_a\right)\cdot \left(\sum_{a}^{N_\eta}\textbf{k}_a\right)\leq \left(\sum_a^{N_\eta}k_a\right)^2\Rightarrow \sum_a^{N_\eta}k_a-p>0,
\end{equation}
which is just a straightforward generalisation of the triangle inequality.} and so exchanging the order of the sum and the integral is invalid. However, the term in which $p\eta$ enters with a negative sign in the exponential isn't taken to the infinite past due to the heaviside theta function. Therefore, the early time limit of this exponential is trivial and we are justified in performing this operation. The effect of taking this early time to be finite is to change the $i\pi$ in \cref{eq:Gamman} to $-Ei(ik\eta')$. We have already argued that the coefficient of this term vanishes and so all our conclusions are still valid. 

This observation also protects the case of linear mixing between massless particles\cite{pimentel2022boostless,jazayeri2022slow} from logarithmic divergences. In these theories it is necessary to calculate a mixed propagator which is given (in the absence of time derivatives) by
\begin{align}
    \tilde{K}_k(\eta')&=\int_{-\infty}^{\eta_0} \frac{d\eta}{\eta^{d+1}}\eta^{2m}F(\textbf{k})K_k^h(\eta) G_k^{h'}(\eta,\eta')\\&=e^{-ic_{h'}k\eta'}P_k^*(\eta')\int_{-\infty}^{\eta'} \frac{d\eta}{\eta^{d+1}}\eta^{2m}F(\textbf{k})\frac{iP_k(\eta)P_k(\eta)}{2k^d(d-2)!!^2}e^{ik\eta(c_{h'}+c_h)}\\&+e^{ic_{h'}k\eta'}P_k(\eta')\int_{\eta'}^{\eta_0} \frac{d\eta}{\eta^{d+1}}\eta^{2m}F(\textbf{k})\frac{iP_k^*(\eta)P_k(\eta)}{2k^d(d-2)!!^2}e^{ik\eta(c_h-c_{h'})} \\&-e^{ic_{h'}k\eta'}P_k(\eta')\int_{-\infty}^{\eta_0}\frac{d\eta}{\eta^{d+1}}\eta^{2m}F(\textbf{k})\frac{iP_k(\eta)P_k(\eta)}{2k^d(d-2)!!^2}e^{ik\eta(c_{h'}+c_h)}.
\end{align}
Each of these terms has a form to which all of our previous arguments apply. Therefore, this mixed propagator can be expressed as a polynomial multiplied by an exponential with a Taylor expansion containing no odd powers of $\eta$ less than $\eta^d$. If we allow time derivatives to act on only the bulk-boundary propagator then nothing changes in this argument. We can still ensure that no time derivatives act on the bulk-bulk propagator. This follows from an argument proceeding exactly as in \cref{eq:Gderiv} except the product of other propagators is absent and we will be left with a term that looks identical to a second spatial derivative. An unfortunate point to note is that the linear mixing considered in both \cite{pimentel2022boostless} and \cite{jazayeri2022slow} is not exclusively built from terms with at least two derivatives. In addition to terms with additional derivatives they consider interactions of the form $\dot{\phi}\sigma$ and this analysis will not apply to such theories. Indeed, in \cite{pimentel2022boostless}, it was shown that this single derivative case had a logarithmic divergence but that all terms with additional derivatives did not, in agreement with our result.

Simple exchange diagrams for $3+1$ dimensional de Sitter as well known in the literature so the cancellation of their divergences will come as little surprise. Therefore, as an illustrative example consider the $4$ point exchange diagram from the interaction ${\phi'}^2\phi$ for a massless scalar in $5+1$ dimensions. We look exclusively at the diagram with all derivatives on external lines for simplicity. Computing the contribution for $\eta>\eta'$ gives
\begin{align}
    \psi_4\supset\lim_{\eta_0\rightarrow 0}i\int_{-\infty}^{\eta_0}\frac{d\eta}{\eta^4}\int_{-\infty}^\eta \frac{d\eta'}{{\eta'}^4}K_{k_1}'(\eta')K_{k_2}'(\eta')K_{k_3}'(\eta)K_{k_4}'(\eta)\left(\phi^-_S(\eta)-\phi^+_S(\eta)\frac{\phi^-_S(\eta_0)}{\phi^+_S(\eta_0)}\right)\phi^+_S(\eta').
\end{align}
In this expression we have introduced the modefunctions $\phi^\pm_k$ for this field which is
\begin{equation}
    \phi^\pm_k(\eta)=\frac{3}{\sqrt{2k^5}}\left(1\mp ik\eta-\frac{1}{3}k^2\eta^2\right)e^{\pm ik \eta}\rightarrow K_k(\eta)=\frac{\phi^+_k(\eta)}{\phi^+_k(\eta_0)}.
\end{equation}
Performing just this part of the integral gives a series of rational terms and a logarithmic divergence
\begin{equation}
    \psi_4\supset \frac{k_1^2k_2^2k_3^2k_4^2}{36S^5}(k_1^2+k_2^2-k_3^2-k_4^2)\log\left(\frac{k_1+k_2+k_3+k_4+2s}{k_1+k_2+k_3+k_4}\right).
\end{equation}
Just as we argued below \cref{eq:exchange} this term is free from logarithmic divergences in time but it does have this logarithmic divergence in the energies. 

This term individually avoids the arguments outlined above because, in the exponential ansatz, it arises from the combination of two terms with different exponents.  However, the contribution from $\eta<\eta'$ can be obtained by exchanging $k_1\leftrightarrow k_3$ and $k_2\leftrightarrow k_4$, exactly cancelling this divergence. This was guaranteed as the logarithm generated from integrals over $e^{i(k_1+k_2+k_3+k_4+2s)\eta}$ can only be seen in this truncated expression. Looking at \cref{eq:Green} we recognise that this term is the same for both $\eta>\eta'$ and $\eta<\eta'$ and so is separable into two contact diagrams neither of which permit such a divergence. This returns us to the situation where there is only a single exponent that could give logarithmic divergences in energy which is forbidden in the absence of such divergences in time.


\subsection{Polynomial Time Divergences}
In addition to forbidding logarithmic time divergences we can also draw some conclusions about the remaining polynomial time divergences. All negative powers of $\eta$ in the integrand are generated by products arising from the first $d$ terms of the sum,
\begin{equation}
    \sum_{n=0}^{\infty}\frac{ \Gamma\left(1-\frac{d}{2}\right)(c_hk\eta)^{2n}}{(-4)^n\Gamma(1+n)\Gamma\left(1-\frac{d}{2}+n\right)},
\end{equation}
or its derivatives. All coefficients in this sum are real and so for parity even interactions their contribution to the wavefunction coefficients, which is built from $iS$, will be imaginary. Similarly, parity odd interactions will have real divergent contributions. Therefore, in both parity odd and even cases the correlation functions can have no contribution from these divergent terms. This is still true for exchange diagrams as the contribution to the integral in \cref{eq:PolyGreen} is real. So, having integrated out all the vertices, the coefficients of the polynomial in the final time integral will be real for parity even interactions (and imaginary for parity odd ones). Thus the divergent contributions to the action will be absent from the correlation functions. 

This result was also noticed in\cite{anninos2015late} and, via an analytic continuation to EAdS and was related in \cite{Maldacena2003} to the ability to holographically renormalise,\cite{Skenderis_2002}, the theory with real counter terms. Relating these divergences to holographic renormalisation in this way suggests an interpretation for their source. Holographic renormalisation can be thought of as subtracting off divergences that arise due to the infinite size of the spacetime. Thus, these divergences can be understood as coming from the infinite time and volume in which the fields can interact. In parity even theories the only possible interactions that can lead to physical divergences are thus those that are logarithmically divergent. As we have seen, such divergences are absent in two, or higher, derivative theories. The requirement for scale invariance means that such interactions become less common as the spacetime expands. So, in spite of the interactions carrying on forever, they become so rare that there are no divergences. 


\section{Extensions and Conclusion}\label{sec:Conclusion}
We have established that there is no way to generate logarithmic divergences at tree level for scale-invariant, parity-even theories involving interactions with at least two derivatives in even spacetime dimensions. What happens when we break these assumptions? In parity odd theories we are allowed odd numbers of spatial derivatives and so our arguments will fail in general. The time integrand can have negative odd powers of $\eta$ provided the number of derivatives is less than $d+1$. Unfortunately, this spoils our argument and so it is possible for the time integral to generate logarithmic divergences as was shown in \cite{cabass2022bootstrapping}.  However, having sufficiently many derivatives (at least $d+1$) acting on each term will remove all negative powers of $\eta$ in the integrand and prevent logarithmic divergences. Our arguments fail even more catastrophically in odd spacetime dimension due to the presence of logarithms in the propagators. This is not saved in a straightforward way by introducing additional derivatives. 

For theories with fewer than two derivatives the arguments straightforwardly break down as there will, generically, be $\frac{1}{\eta}$ terms in the integrand. This is already known and is seen, for example, in the case of $\phi^3$ interactions\cite{falk1992angular}. In the case of massive fields, the propagators that we generate from the Frobenius ansatz are made out of non-integer powers of $\eta$ and so this analysis of divergences is invalid, with the exception of a finite number of specific masses that have odd half integer $\nu$. Likewise, in the exponential ansatz, the function multiplying the exponential will not terminate and so we cannot write it as a polynomial. Therefore, we need to worry about the exchange of the sum and the integral. This is also what will happen if we alter the background theory. For example by allowing the mass or speed of sound to vary in time or if we introduce a parity breaking term in the quadratic action. A potential extension to this work would be to consider the implications for fermionic fields which were ignored in this work. It may also be constructive to explore the behaviour of loop diagrams. However, logarithmic divergences are expected to be completely generic in such cases.

In conclusion, all tree level wavefunction coefficients of scale invariant, massless, integer spin fields, involving parity even interactions with at least two derivatives in even spacetime dimension contain no logarithmic divergences in time or momentum and so are rational functions of the momenta plus potentially some polynomial time divergences in the limit. Furthermore, any such time divergences are guaranteed to be imaginary and thus do not affect correlation functions. These assumptions, whilst restrictive, are not just made for the sake of convenience, any massive fields present during the early universe are expected to decay and so masslessness is required to make predictions for observations. Furthermore, in spite of the insistence on scale invarance restricting us to a de Sitter background we allow boost breaking both through arbitrary sound speeds and non de Sitter invariant interactions.

This is an important ingredient in the bootstrap program as it heavily constrains the allowed forms of wavefunction coefficients by reducing the space of allowed functions that they can take to a polynomial basis with an order fixed by the number of derivatives and the requirement of scale invariance. This result breaks down in the presence of massive fields and so a discovery of this sort of divergence in the data would be a powerful result for the cosmological collider which seeks to understand the spectrum of massive states in the early universe.

\section*{Acknowledgements}
I would like to thank Enrico Pajer for initial collaboration and helpful comments about the draft, James Bonifacio and David Stefanyszyn for useful discussions over the course of the work and Aaron Hillman and Dong-Gang Wang for their comments on the draft. The author is supported by jointly by the Science and Technology Facilities Council through a postgraduate studentship and the Cambridge Trust Vice Chancellor's Award.

\bibliographystyle{JHEP}
\bibliography{ref}

\end{document}